\documentclass[twoside,10pt,a4paper]{newFNLstyle}
\usepackage{graphics}
\usepackage{cite}

\begin{document}

\volnumpagesyear{1}{?}{L???--L???}{2002}
\dates{30 May 2002}{30 May 2002}{30 May 2002}

\title{THE PHYSICAL BASIS FOR PARRONDO'S GAMES }

\authorsone{ANDREW ALLISON and DEREK ABBOTT }

\affiliationone{Centre for Biomedical Engineering (CBME) and EEE
  Dept., }

\mailingone{Adelaide University, SA 5005, Australia. \\
  {\it Email:}~aallison@eleceng.adelaide.edu.au,~dabbott@eleceng.adelaide.edu.au}

\maketitle

\markboth{Allison \& Abbott}{Stochastic Resonance in a Brownian Ratchet}

\pagestyle{myheadings}


\begin{abstract}

% statement of problem and context}
Several authors 
\cite{harmer_1999,Harmer_1999b,harmer_2000,harmer_2001} 
have implied that
the original inspiration for Parrondo's games was a
physical system called a ``flashing Brownian ratchet
\cite{Doering_1995,doering_1998}''  The relationship seems to be intuitively
clear but, surprisingly, has not yet been established with rigor.

% statement of scope of the paper
The dynamics of a flashing Brownian ratchet can be described using a
partial differential equation called the Fokker-Planck equation
\cite{risken_1988}, that describes the probability density, of finding 
a particle at a certain place and time, under the influence of diffusion 
and externally applied fields. In this paper, we apply standard finite-difference
methods of numerical analysis \cite{lapidus_1962,scheid_1968,press_1988} to
the Fokker-Planck equation. We derive a set of finite difference equations
and show that they have the same form as Parrondo's games.
This justifies the claim that Parrondo's games are a discrete-time, discrete-space
version of a flashing Brownian ratchet. Parrondo's games, are in effect,
a particular way of sampling a Fokker-Planck equation. Our difference equations
are a natural and physically motivated generalisation of Parrondo's games.
We refer to some well established theorems of numerical analysis to suggest conditions
under which the solutions to the difference equations and partial differential
equations would converge.

% indication of possible future directions
The diffusion operator, implicitly assumed in Parrondo's original
games, reduces to the Schmidt formula for the integration of the
diffusion equation. There is actually an infinite continuum of
possible diffusion operators. The Schmidt formula is at one extreme of
the feasible range.  We suggest that an operator in the middle of the
feasible range, with half-period binomial weightings, would be a
better representation of the underlying physics.
% response to the reviewer's comments on the ``MEMS Ratchet.''

%possible application
Physical Brownian ratchets have been constructed and have worked
\cite{faucheux_1995,slater_1997,ertas_1998,duke_1998,bader_1999}. It
is hoped that the finite element method presented here will be useful
in the simulation and design of flashing Brownian ratchets.
 
\end{abstract}

\section{The Fokker-Planck equation}

One of the classical problems of statistical physics, and physical
chemistry, is to find a macroscopic statistical description for the
diffusion of a dissolved molecule or ion in a uniform fluid
solvent. The microscopic state of such a system has very many degrees
of freedom, possibly even more than an Avogadro number of degrees of
freedom. This gives rise to an equally large number of coupled
equations of motion.  It is completely impractical to solve such a
large system with rigor. We must abandon the idea of an exact
solution.  We are forced to use a statistical description, where we can
only describe the probability of certain events.

We denote the probability of finding a Brownian particle at a certain
point on space, $z$, and time, $t$, by $p = p\left( z, t \right)$.
The time-evolution of $p\left( z, t \right)$ is governed by a partial differential equation called
the Fokker-Planck equation:
\begin{equation}
   \frac{\partial^{2}}{\partial z^2}  \left( D\left( z,t \right)      p \left( z, t \right) \right)
   - \frac{\partial}{\partial z}      \left( \alpha \left( z, t \right) p \left( z, t \right) \right)
   -   \frac{\partial }{\partial t}  p \left( z, t \right)
   = 0~.
   \label{eq:fokker_planck}
\end{equation}

The functions $\alpha \left( z, t \right)$ and $D\left( z,t \right)$
are referred to as the infinitesimal first and second moments of
diffusion. In practice, the infinitesimal second moment does
sometimes depend on concentration of the solute, 
$p\left( z, t \right)$, but is usually regarded as constant and is called the
``Fick's law constant.'' A typical value (for a hydrated sodium ion in water)
 would be of the order $D \approx 1.3 \times 10^{-9}~ \mbox{m}^{2} \mbox{s}^{-1}$.
The infinitesimal first moment depends on the magnitude of externally imposed forces and on  the mobility of the Brownian particle which is given by
\begin{equation}
  u = \frac{Z_e}{6 \pi \eta a}
  \label{eq:mobility}
\end{equation}
where $Z_e$ is the electrical charge on the particle,
$\eta$ is the kinematic viscosity of the solvent
and $a$ is the effective radius of the particle.
A typical value for the mobility (of a hydrated sodium ion in water) would be
$u \approx 51.9 \times 10^{-9}~ \mbox{m}^{2} \mbox{s}^{-1} \mbox{volt}^{-1}$. 
Further descriptions and numerical data
may be found in books on physical chemistry and statistical physics 
\cite{bird_1960,reif_1965,atkins_1978,cussler_1997}. 
If we apply an electrical potential, or voltage,
of $V\left( z, t \right)$ then the infinitesimal first moment is given by
\begin{equation}
   \alpha \left( z , t \right) = - u \frac{\partial}{\partial z} V\left( z, t \right)~.
   \label{eq:infinitesimal_first_moment}
\end{equation}
The theory behind Equations \ref{eq:mobility} and \ref{eq:infinitesimal_first_moment}
is due to Stokes and Einstein \cite{einstein_1956}.
More information about the methods of solution and the applications of
the Fokker-Planck equation can be found in Risken \cite{risken_1988}.

When we take into account the functional forms of $D$ and $\alpha$ then 
we can rewrite the Fokker-Planck equation as:
\begin{equation}
   D \frac{\partial^{2} p }{\partial z^2}
   - \frac{\partial \alpha }{\partial z } p
   - \alpha \frac{\partial p }{\partial z}
   -   \frac{\partial p }{\partial t}
   = 0~.
   \label{eq:expanded_fokker_planck}
\end{equation}
This is the form of the Fokker-Plank equation which we will sample at
regular intervals in time and space, to yield finite difference
equations.

\section{Finite difference approximation}

Many Partial Differential Equations, or PDEs, including Equation
\ref{eq:expanded_fokker_planck}, can be very difficult to solve
analytically. One well established approach to this
problem is to sample possible solutions to a PDE
at regular intervals, called mesh points \cite{lapidus_1962}.  The true solution is
approximated locally by a collocating polynomial. The values of the
derivatives of the true solution are approximated by the
corresponding derivatives of the collocating polynomial.

We can define local coordinates, expanded locally about a point
$\left( z_0 , t_0 \right)$ we can map points between a real space
$\left( z , t \right)$ and an integer or discrete space 
$\left( i , j \right)$. Time, $t$, and position, $z$, are modelled by real numbers,
$t, z \in \mathcal{R}$ and the corresponding sampled position, $i$, and
sampled time, $j$, are modelled by integers 
$i, j \in \mathcal{Z}$. We sample the space using a simple linear relationship
\begin{equation}
   \left(z , t \right) =  \left(z_0 + i \lambda , t_0 + j \tau \right)
   \label{eq:sample}
\end{equation}
where $\lambda$ is the sampling length and $\tau$ is the sampling time.

In order to map Equation \ref{eq:expanded_fokker_planck} into discrete
space, we need to make suitable finite difference approximations to
the partial derivatives. The notation is greatly simplified if we define a family of difference operators:
\begin{equation}
   \Delta_{i,j} = p \left(z_0 + i \lambda , t_0 + j \tau \right) - p \left(z_0 , t_0 \right).
   \label{eq:difference}
\end{equation}
In principle, this is a doubly infinite family of operators but in
practice we only use a small finite subset of these operators. This is
determined by our choice of sampling points. This choice is not unique
and is not trivial. The set of sampling points is called a
``computational molecule \cite{lapidus_1962}.'' Some choices lead to over-determined sets
of equation with no solution. Some other choices lead to
under-determined sets of equations with infinitely many solutions.  We
chose a computational molecule called ``Explicit'' computation with the
following sample points:
$\left( i, j \right) \in \left\{ (0,0) , (-1,-1) , (0,-1) , (+1,-1) \right\}$.

We also need to make a choice regarding the form of the local
collocating polynomial. This is not unique and inappropriate choices
do not lead to unique solutions. A polynomial which is quadratic in $z$
and linear in $t$ is the simplest feasible choice:
\begin{equation}
   p \left(z,t \right) =  p \left( z_0, t_0 \right)  +
        A_1 \cdot \left( z - z_0 \right)   +
        A_2 \cdot \left( z - z_0 \right)^2 +
        B_1 \cdot \left( t - t_0 \right) 
   \label{eq:polynomial}
\end{equation}
where $A_1$, $A_2$ and $B_1$ are the real coefficients of the
polynomial.
Equations \ref{eq:sample}, \ref{eq:difference} and \ref{eq:polynomial}
imply a simple system of linear equations that can be expressed in matrix form:
\begin{equation}
\left[
\begin{array}{ccc}
   -\lambda   & +\lambda^2  & -\tau  \\
   0        &  0         & -\tau  \\
   +\lambda   &  +\lambda^2  & -\tau
\end{array}
\right]
\left[
\begin{array}{c}
     A_1 \\
     A_2 \\
     B_1
\end{array}
\right]
=
   \left[
   \begin{array}{c}
       \Delta_{-1,-1} \\
       \Delta_{0,-1} \\
       \Delta_{+1,-1}
   \end{array}
   \right]~.
\label{eq:finite_difference_conversion}
\end{equation}
These can be solved algebraically, using Cramer's method to obtain expressions
for $A_1$, $A_2$ and $B_1$:
\begin{equation}
   A_1 = \frac{p \left( z_0 + \lambda , t_0 - \tau \right) - p \left( z_0 - \lambda , t_0 - \tau \right)}
         {2 \lambda}
   \label{eq:A_1}
\end{equation}
and
\begin{equation}
   A_2 = \frac{   p \left( z_0 - \lambda , t_0 - \tau \right) 
             -2 p \left( z_0         , t_0 - \tau \right)
             +  p \left( z_0 + \lambda , t_0 - \tau \right) }
         {2 \lambda^2}
   \label{eq:A_2}
\end{equation}
and 
\begin{equation}
   B_1 = \frac{ p \left( z_0 , t_0 \right) - p \left( z_0 , t_0 - \tau \right) }
            { \tau }~.
   \label{eq:B_1}
\end{equation}
These are all intuitively reasonable approximations but their choice
is not arbitrary.  Equations \ref{eq:A_1}, \ref{eq:A_2}, \ref{eq:B_1}
form a complete and consistent set. We could not change one without
adjusting the others.  We can evaluate the derivatives of Equation
\ref{eq:polynomial}  to obtain a complete and consistent set of
finite difference approximations for the partial derivatives:
\begin{equation}
   \frac{\partial p }{\partial z} =   
   A_1 = \frac{p \left( z_0 + \lambda , t_0 - \tau \right) - p \left( z_0 - \lambda , t_0 - \tau \right)}
         {2 \lambda}
   \label{eq:dpdz}
\end{equation}
and
\begin{equation}
   \frac{\partial^2 p }{\partial z^2} =
   2 A_2 = \frac{   p \left( z_0 - \lambda , t_0 - \tau \right) 
             -2 p \left( z_0         , t_0 - \tau \right)
             +  p \left( z_0 + \lambda , t_0 - \tau \right) }
         { \lambda^2}
   \label{eq:d2pdz2}
\end{equation}
and
\begin{equation}
   \frac{\partial p }{\partial t} = 
   B_1 = \frac{ p \left( z_0 , t_0 \right) - p \left( z_0 , t_0 - \tau \right) }
            { \tau }~.
   \label{eq:dpdt}
\end{equation}
We can apply the same procedure to $\alpha \left( z, t \right)$ to obtain
\begin{equation}
   \frac{\partial \alpha }{\partial z} =   
   A_1 = \frac{\alpha \left( z_0 + \lambda , t_0 - \tau \right) 
             - \alpha \left( z_0 - \lambda , t_0 - \tau \right)}
         {2 \lambda}~.
   \label{eq:dalphadz}
\end{equation}

Equations \ref{eq:dpdz}, \ref{eq:d2pdz2}, \ref{eq:dpdt} and \ref{eq:dalphadz}
can be substituted into Equation \ref{eq:expanded_fokker_planck} to yield the required
finite partial difference equation:
\begin{equation}
    p \left(z_0 , t_0 \right)
    = a_{-1} \cdot p \left(z_0 - \lambda , t_0 - \tau \right) +
      a_{0}  \cdot p \left(z_0         , t_0 - \tau \right) +
      a_{+1} \cdot p \left(z_0 + \lambda , t_0 - \tau \right)
   \label{eq:partial_difference_equation}
\end{equation}
where
\begin{equation}
   a_{-1} = \frac
     {\frac{D \tau}{\lambda^2} + \frac{\alpha \left( z_0 , t_0 \right) \tau}{2 \lambda} }
     {\frac{\alpha \left( z_0 + \lambda , t_0 - \tau \right) -
          \alpha \left( z_0 - \lambda , t_0 - \tau \right) }{2 \lambda} \tau + 1 }
   \label{eq:am1}
\end{equation}
and
\begin{equation}
   a_{0} = \frac
     {  -2 \frac{ D \tau}{\lambda^2} + 1 }
     {\frac{\alpha \left( z_0 + \lambda , t_0 - \tau \right) -
          \alpha \left( z_0 - \lambda , t_0 - \tau \right) }{2 \lambda} \tau + 1 }
   \label{eq:a0}
\end{equation}
and
\begin{equation}
   a_{+1} = \frac
     {\frac{D \tau}{\lambda^2} - \frac{\alpha \left( z_0 , t_0 \right) \tau}{2 \lambda} }
     {\frac{\alpha \left( z_0 + \lambda , t_0 - \tau \right) -
          \alpha \left( z_0 - \lambda , t_0 - \tau \right) }{2 \lambda} \tau + 1 }~.
   \label{eq:ap1}
\end{equation}
We can overload the arguments of $p$ and write them in terms of the
discrete space $\left( i , j \right)$ using the mapping defined in Equation \ref{eq:sample}.
\begin{equation}
    p_{i  , j }
    = a_{-1} \cdot p_{i -1  , j - 1 } +
      a_{0}  \cdot p_{i     , j - 1 } +
      a_{+1} \cdot p_{i + 1 , j - 1 }~.
   \label{eq:integer_partial_difference_equation}
\end{equation}
The meaning of the arguments should be clear from the context and from
the use of subscript notation, $p_{i,j}$, rather than function
notation, $p(z,t)$.  Equation
\ref{eq:integer_partial_difference_equation} is precisely the form
required for Parrondo's games.

\section{Parrondo's games}

In the original formulation,
the conditional probabilities of winning or losing depend on the state, $i$, 
of capital but not on any other information about the past history of the games:
\begin{itemize}
\item~Game A is a toss of a biased coin:~
\begin{equation}
   p_{win} = \frac{1}{2} - \epsilon
   \label{eq:game_A}
\end{equation}
where $\epsilon$ is an adverse external bias that the game has to
``overcome''. This bias, $\epsilon$, is typically a small number such as 
$\epsilon = 1/200$, for example \cite{harmer_1999,harmer_2000}.

\vspace{0.167in}
\item~Game B depends on the capital, $i$:~
   \begin{description}
      \item[If $(i \bmod 3) = 0$],~then the odds are unfavorable.
         \begin{equation}
            p_{win} = \frac{1}{10} - \epsilon
            \label{eq:bad_game_B}
         \end{equation} 
      \item[If $(i \bmod 3) \neq 0$],~then the odds are favorable.
         \begin{equation}
            p_{win} = \frac{3}{4} - \epsilon~.
            \label{eq:good_game_B}
         \end{equation}
   \end{description}
\end{itemize}
It is straightforward to simulate a randomized sequence of these games
on a computer using a very simple algorithm \cite{harmer_2001}.

\subsection{Game A as a partial difference equation}

We can write the requirements for game A in the form of Equation
\ref{eq:integer_partial_difference_equation}.
\begin{equation}
    p_{i  , j }
    = \left( \frac{1}{2} - \epsilon \right) \cdot p_{i -1  , j - 1 } +
      0 \cdot p_{i     , j - 1 } +
      \left(  \frac{1}{2} + \epsilon \right ) \cdot p_{i + 1 , j - 1 }~.
   \label{eq:game_A_partial_difference_equation}
\end{equation}
This implies a constraint that $a_0 = 0$ which implies that 
${D \tau}/{\lambda^2} = 1/2$ which defines the relative scales of
$\lambda$ and $\tau$ so we can give it a special name:
\begin{equation}
   \beta = \frac{D \tau}{\lambda^2}~.
   \label{eq:beta}
\end{equation}

The constraints on $a_{-1}$ and $a_{+1}$ imply a value for Parrondo's ``$\epsilon$''
parameter:
\begin{equation}
   \epsilon = \left\{ \frac{\lambda}{4 D} \right\} \alpha \left( z_0 , t_0 \right)
   \label{eq:epsilon}
\end{equation}
which can be related back to an externally imposed electric field, 
$E = - \partial V / \partial z$ using equations
\ref{eq:mobility} and \ref{eq:infinitesimal_first_moment}:
\begin{equation}
  \epsilon = \left( \frac{\lambda}{4 D} \right) 
          \left( \frac{Z_e}{6 \pi \eta a} \right) 
          \left( - \frac{\partial V}{\partial z} \right)~.
  \label{eq:physical_epsilon}
\end{equation}
The small bias, $\epsilon$, is proportional to the applied external
field which justifies Parrondo's original intuition.

\subsection{Game B as a partial difference equation}

There is still zero probability of remaining in the same state
which implies a constraint that $a_0 = 0$ which implies that we still have
the same scale, $\beta = \frac{1}{2}$. If we are in state $i$ then we can denote
the probability of winning by \\ 
$q_i  = P\left( \mbox{win} | \mbox{initial position is} ~i \right)$.
We can write the difference equations for game B in the form:
\begin{equation}
    p_{i  , j }
    = q_{i-1}              \cdot p_{i -1  , j - 1 } +
      0                    \cdot p_{i     , j - 1 } +
      \left( 1- q_{i+1} \right) \cdot p_{i + 1 , j - 1 }~.
   \label{eq:game_B_partial_difference_equation}
\end{equation}
which, together with Equations \ref{eq:am1}, \ref{eq:a0} and \ref{eq:ap1},
gives
\begin{equation}
   \frac{q_{i-1}}{1 - q_{i+1}}
                     = \frac{a_{-1}}{a_{+1}}
                     = \frac{ 1 + \frac{\lambda}{2 D \tau} \alpha_{i,j} }
                          { 1 - \frac{\lambda}{2 D \tau} \alpha_{i,j} }
   \label{eq:q_rat}
\end{equation}
which implies that
\begin{eqnarray}
   \alpha_{i,j} = 2 \lambda \beta \frac{q_{i-1} - \left( 1 - q_{i+1} \right) }
                                {q_{i-1} + \left( 1 - q_{i+1} \right)}~.
  \label{eq:physical_alpha}
\end{eqnarray}
This can be combined with Equation \ref{eq:infinitesimal_first_moment}
and then directly integrated to calculate the required voltage profile.
We can approximate the integral with a Riemann sum:
\begin{equation}
   V_i = - \frac{2 \beta}{u} \sum_{k=0}^{i}
         \frac{1 - \left( \frac{1 - q_{k+1}}{q_{k-1}} \right)}
            {1 + \left( \frac{1 - q_{k+1}}{q_{k-1}} \right)}
   \label{eq:integer_voltage}
\end{equation}
so we can construct the required voltage profile for the ratchet which
means that, given the values of $q_i$, it is possible to construct a
physical Brownian ratchet that has a finite difference approximation
which is identical with Parrondo's games. We can conclude that
Parrondo's games are literally a finite element model of a flashing
Brownian ratchet.

We note that game B, as defined here, is quite general and actually includes
game A as a special case.

\subsection{Conditions for convergence of the solution}
We would like to think that as long as $\beta = D \tau / \lambda^2 $
is preserved then the solution to the finite partial difference
equation
\ref{eq:integer_partial_difference_equation}
would converge to the true solution of the partial differential
equation
\ref{eq:expanded_fokker_planck}, as the mesh size, $\lambda$ goes to zero.
Fortunately, there is a theorem due to O'Brien, Hyman and Kaplan
\cite{obrien_1951}
which establishes that the numerical integration of a parabolic
PDE, in explicit form, will converge to the correct solution as
$\lambda \rightarrow 0$ and $\tau \rightarrow 0$ provided $\beta \leq \frac{1}{2}$.
Similar results may also be found in standard texts on 
numerical analysis \cite{lapidus_1962,scheid_1968,press_1988}.

We see that Parrondo's choice of diffusion operator, with 
$\beta = \frac{1}{2}$ is at the very edge of the stable region.

\subsection{An appropriate choice of scale}

There is a possible range of values for $\beta$. As $\beta \rightarrow
0$ we require the time step $\tau \rightarrow 0$ which means that the
number of time steps required to simulate a given time interval, $T$,
increases without bound $N_{\mbox{steps}} = T/\tau \rightarrow \infty$. It is
computationally infeasible to perform simulations with very small
values of $\beta$.  On the other hand, the value of $\beta = 1/2$
implied in Parrondo's games is at the very limit of stability. In fact,
the presence of small roundoff errors in the arithmetic could cause the
the discrete simulation to diverge significantly from the continuous
solution.

We propose that choosing $\beta = 1/4$, in the middle of the feasible
range, is most appropriate. If we consider the case of pure diffusion, with
$\alpha = 0$, then Equation \ref{eq:integer_partial_difference_equation} reduces to
\begin{equation}
    p_{i  , j }
    = \beta \cdot p_{i -1  , j - 1 } +
      \left( 1 - 2 \beta \right)  \cdot p_{i     , j - 1 } +
      \beta \cdot p_{i + 1 , j - 1 }
   \label{eq:diffusion_difference_equation}
\end{equation}
and if we choose $\beta = 1/4$ then this reduces to
\begin{equation}
    p_{i  , j }
    = \frac{ 1 \cdot p_{i -1  , j - 1 } +
           2 \cdot p_{i     , j - 1 } +
           1 \cdot p_{i + 1 , j - 1 }}{4}
   \label{eq:binomial_difference_equation}
\end{equation}
which is the same as Pascal's triangle with every second row removed.
The solution to the case where the initial condition is a Kronecker delta function,
$p_{i,0} = \delta_{i,0}$ is easy to calculate:
\begin{equation}
    p_{i  , j }
    = \frac{1}{2^{2j}} \cdot \left(^{2 j}_{j+i} \right)
    = \frac{1}{2^{2j}} \cdot \frac{ \left( 2 j \right)!}{  \left( j + i \right)!  \left( j - i \right)! }
   \label{eq:binomial_solution}
\end{equation}
which is a half period, or double frequency, binomial. We can invoke
the Laplace and De Moivre form of the Central Limit Theorem which
establishes a correspondence between Binomial (or Bernoulli)
distribution and the Gaussian distribution to obtain
\begin{equation}
   p_{i,j} = \frac{1}{ \sqrt{2 \pi \left( \frac{j}{2} \right) }} 
     \exp \left( \frac{ -i^2 }{ j }  \right)~.
   \label{eq:laplace_demoivre}
\end{equation}
This expression is only approximate but is true in the limiting case as
$j \rightarrow \infty$.

In the case where $\alpha=0$; the Fokker Planck Equation \ref{eq:expanded_fokker_planck}
reduces to a diffusion equation:
\begin{equation}
   D  \frac{{\partial}^{2} p}{\partial {z}^{2}} - \frac{\partial p}{\partial t} = 0
   \label{eq:diffusion_1D}
\end{equation}
Einstein's solution to the diffusion equation is a Gaussian
probability density function:
\begin{equation}
   p \left( z , t \right) = \frac{1}{\sigma \sqrt{2 \pi}} 
     \exp \left( \frac{ -z^2 }{ 2 \sigma ^ 2}  \right)
   \label{eq:Gaussian}
\end{equation}
where the variance, $\sigma^2$, is a linear function of time:
\begin{equation}
   \sigma ^2 = 2 D t~.
   \label{eq:variance}
\end{equation}
It is possible to verify that this is a solution by direct substitution:
\begin{eqnarray}
   D  \frac{{\partial}^{2} p}{\partial {z}^{2}} & = & \frac{\partial p}{\partial t} \\
  ~  & = & \left( \frac{-1}{2 t} \right) \cdot \left( 1 - \left( \frac{z}{\sigma} \right)^2 \right) \cdot p \left( z , t \right)~.
   \label{eq:verification}
\end{eqnarray}
If we sample this solution in Equation \ref{eq:Gaussian} 
using the mapping in Equation \ref{eq:sample} 
then we obtain Equation \ref{eq:laplace_demoivre} again. 
This is an exact result. We conclude that the choice of $\beta
= 1/2$ is very appropriate for the solution to the diffusion
equation. We suggest that this would also be true for the
Fokker-Planck equation, in the case where $\alpha$ is ``small.''  The
appropriate choice of $\beta$, given arbitrarily large, or rapidly
varying, $\alpha$ is still an unsolved problem.
In general, we would expect that much smaller values, 
$\beta \rightarrow 0$, 
would be needed to accommodate more extreme choices of alpha.

\subsection{An example of a simulation}

\begin{figure}[\h!] 
\centering{\resizebox{3.3in}{!}{\includegraphics{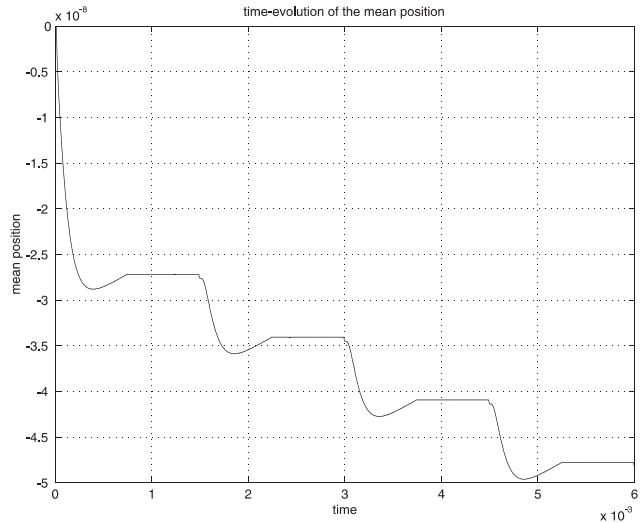}}}
\caption{ \small  
      Time-evolution of the mean of the distribution $p(z,t)$. When
      the field is asserted, the mean position of the particles moves
      in a generally ``downward'' direction. There is some relaxation
      towards the end of that part of the cycle. When the field is
      turned off, the mean remains constant although diffusion causes
      the field to spread. The total shift in mean position of this
      ratchet is very modest, about $0.005~\mu$m, compared with the
      spacing between the teeth of the ratchet, of $2.0~\mu$m. Part
      of the motivation of this work is to optimise the transport
      effect of the Brownian ratchet, subject to constraints.
      \normalsize}
\label{fig:simulation} 
\end{figure}

We simulated a physically reasonable ratchet with a moderately large
modulo value, $M=8$. (The value for the original Parrondo's games was
$M=3$.) We used the value of $\beta = 1/4$. The simulation was based
on a direct implementation of Equation
\ref{eq:integer_partial_difference_equation} in Matlab. 
We chose a sampling time of $\tau = 12~\mu\mbox{s}$ and a sampling
distance of $\lambda \approx 0.25 \mu\mbox{m}$.  The result is shown
in Figure \ref{fig:simulation}, where we indicate how the expected
position of a particle can move within a Brownian flashing ratchet
during four cycles of the modulating field. We can see a steady drift
of the mean position of the particle in response to the ratchet
action.
%-------------
%-------------
This simulation includes a total of 500 time samples. Note that the
average rate of transport quickly settles down to a steady value, even
after only four cycles of the ratchet.

\section{Conclusions}

We acknowledge the similar, but independent, work of Heath
\cite{heath_2002} et al. The focus of our paper is different.
We seek to establish the physical, and mathematical, basis of
Parrondo's games and to derive a practical numerical technique for
simulation.

We conclude that Parrondo's games {\it are} a valid finite-element
simulation of a flashing Brownian ratchet, which justifies Parrondo's
original intuition. We have established that Parrondo's ``$\epsilon$''
parameter is a reasonable way to simulate a gradual externally imposed
electric field, or voltage gradient. We have established that
Parrondo's implied choice of the $\beta$ parameter does lead to a
stable simulation but we suggest that the choice of $\beta = 1/4$ is
more appropriate from a mathematical point of view.

Finally, we have generalised Parrondo's games, in the form of a set of
finite difference equations
\ref{eq:integer_partial_difference_equation} and we have shown that
these can be implemented on a computer.

% \section*{Acknowledgments}
% The authors like to acknowledge funding from the Australian Research
% council and the GTECH Corporation Australia.

\end{document}